\documentstyle[12pt]{article}

\begin{document}
\baselineskip=1.5\normalbaselineskip
\begin{center}
{\Large\bf Transition to nonchaotic behavior in a Brownian-type 
motion}\\[\baselineskip]
{B. Kaulakys$^*$ and G. Vektaris$^\dagger$}\\[\baselineskip]
{Institute of Theoretical Physics and Astronomy,}\\
{A. Go\v stauto 12, 2600 Vilnius, Lithuania}\\[2\baselineskip]

\parbox{5in}{\baselineskip=1.5\normalbaselineskip
A theoretical and numerical analysis of the 
transition from chaotic to nonchaotic behavior in an ensemble of particles
with different initial conditions which move according to Newton's equations
in a bounding potential and are driven by an identical sequence of random
forces (see S. Fahy and D. R. Hamann, Phys. Rev. Lett. {\bf 69}, 761 (1992))
is presented. The threshold values of the parameters for transition from
chaotic to nonchaotic behavior are defined on the basis of the map for distances
between the particles and differences of velocity. Numerical analysis is
fulfilled for one-dimensional Duffing $V(x)=x^4-x^2$ and $V(x)=x^4$
potentials.}\\[1.5\baselineskip]
\end{center}

PACS number(s): 05.40.+j, 05.45.+b

\centerline{}

Recently an interesting transition from chaotic to nonchaotic behavior
in randomly driven systems has been discovered [1]. When an ensemble of
bounded in a fixed external potential particles with different initial
conditions are driven by an identical sequence of random forces, the
ensemble of trajectories may become identical at long times. Fahy and
Hamann [1] considered a particle of mass $m$ moving according to Newton's
equations in a potential $V(x)$, except that at regular time intervals
$\tau$ the particle is stopped and its velocity is reset to random value
chosen from a Maxwell distribution with temperature $T$. It should be
stressed that for every particle of the ensemble it was given an identical,
randomly chosen velocity at the start of each step of time length
$\tau$. This motion is in many respects similar to Brownian motion of the
particles at a temperature $T$ and is frequently used in Monte Carlo
simulations [2]. However, if the time interval $\tau$ between stops is lower
than a threshold value $\tau_c$, the final trajectories of the particles are
independent on the initial conditions; all trajectories become point by
point identical in time. Although the trajectory is highly erratic and
random, the system is not chaotic.

The similar effect may also be observed in a more general and realistic (from
the physical point of view) case, i.e., when mixing at time intervals $\tau$ 
some part $\alpha$ of the old velocity ${\bf v}^{old}$ with random velocity
${\bf v}^{ran}$ to get a new starting velocity
${\bf v}^{new}=\alpha{\bf v}^{old}+{\bf v}^{ran}$ [3]. Here a threshold value
$\tau_c$ depends on $\alpha$.

Note, that the observed effect resembles a phase transition but does not
depend crucially on the dimension of the space in which the particles move.
Besides its intrinsic interest, this phenomenon has some importance for Monte
Carlo simulations [1] and, probably, can have influence on some processes, 
e.g., clustering of particles.

In the paper [1] except the discovery in the computer simulations of the
transition from chaotic to nonchaotic behavior, the analytical estimation of
the average rate of contraction of the distance between initially close
together identically driven particles (for a one-dimensional potential, very 
short time intervals $\tau$ between stops and $\alpha=0$) is given. 
Note to the paper [4] in which the similar problem was analysed and it was
found that trajectories of a noisy logistic map for various initial conditions
may become point by point identical in time if the noise is sufficiently 
strong and independent of the initial condition. Moreover, using Langevin
equation, it was shown quite generally, that nonlinear systems coupled through
identical strong noises may have the same trajectory in the long time limit.
However, the derivation of dependence of the threshold noise on the parameters
of the system [1] is absent.

It is the purpose of this paper to present a theoretical and numerical analysis
of the transition from chaotic to nonchaotic behavior in randomly driven
one-dimensional systems, i.e. the dependencies of the threshold value $\tau_c$
on the parameters of the problem, including $\alpha$, and convergence rates
of the trajectories. Generalization of the method for two- and three-dimensional
systems is rather straightforward (see below). It should be noted that in the 
numerical calculations
of the mean-square distance between particles we observe the intermittency
route to chaos when $\tau\to\tau_c$: mean-square distance as a function
of time alternates randomly between almost clustered phases and irregular
bursts of the dispersed particles. Moreover, we find a transition from chaotic
to nonchaotic behavior not only for the regular time intervals $\tau$ but for
the randomly changing intervals between stops, i.e. for random $\tau_i$
(where $i$ is the number of the step) as well. Our theoretical analysis is
based on the mapping form of the equations of motion for the distance
between the particles and the difference of the velocity of the particles.

Let us consider a particle of mass $m$ moving in a one-dimensional potential
$V(x)$ which confines particles to a finite region. At a time intervals 
$\tau_i$ the particle is partially stopped and its velocity is reset to a
new starting velocity $v_i=\alpha v_i^{old}+v_i^{ran}$. Between the stops
the particle moves according to Newton's equations
$$
{dx\over{dt}}=v,~~~~~~ {d^2x\over{dt^2}}=-{1\over{m}}{dV\over{dx}}.\eqno(1)
$$
When two particles initially at points $x_0$ and $x'_0$ are started with
velocities $v_0$ and $v'_0$ and are driven by an identical sequence of random
velocities $v_i^{ran}$ at the same time intervals $\tau_i$, coordinates
and velocities of them may accidentally draw closer to one another. The
convergence of the two trajectories to the single final trajectory will depend
on the evolution with a time of the small variancies of the distance
$\Delta x_i=x'_i-x_i$ and velocity $\Delta v_i=v'_i-v_i$. Moreover, we
investigate a transition from chaotic to nonchaotic behavior. Generally, such
a transition may be detected from analysis of behavior of the neighboring
trajectories and it is described by the Lyapunov characteristic exponents and 
$KS$ metric entropy of the flow of trajectories in a given region of phase space
[5-9]. Further we will follow this method. Comparisons of the results of such
analysis with investigations by the direct numerical simulations confirm the
suitability of our theoretical approach.

From formal solutions $x=x(x_i,v_i,t)$ and $v=v(x_i,v_i,t)$ of equations (1)
with initial conditions $x=x_i$ and $v=v_i$ at $t=0$ it follows an equation
for $\Delta x(t)$ and $\Delta v(t)$ at a time moment $t$:
$$
\left({\Delta x(t) \atop{\Delta v(t)}}\right)={\bf T}(\alpha;x_i,v_i,t)
\left({\Delta x_i \atop{\Delta v_i}}\right) \eqno (2)
$$
where the matrix {\bf T} is of the form
$$
{\bf T}=\pmatrix{T_{xx}&\alpha T_{xv}\cr
T_{vx}&\alpha T_{vv}\cr}=
\pmatrix{\displaystyle{\partial x(x_i,v_i,t)\over\partial x_i}&
\displaystyle{\alpha{\partial x(x_i,v_i,t)\over\partial v_i}}\cr
\displaystyle{\partial v(x_i,v_i,t)\over\partial x_i}&
\displaystyle{\alpha{\partial v(x_i,v_i,t)\over\partial v_i}}\cr}.\eqno(3)
$$
Note, that the similar method of investigation is used in the theory of
transition to chaos in classical systems [5,6]. However, the motion in the
form (2) and (3) is represented as the non-area-preserving tangent map, while
classical dynamics of the conservative systems may be represented by the
area -preserving maps.

According to equations (1) and (3) matrix elements $T_{xx}$ and $T_{xv}$
satisfy the equation
$$
{d^2T_x\over dt^2}=-{1\over m}{d^2V\over dx^2}\Biggl|_{x=x(x_i,v_i,t)}T_x
\eqno(4)
$$
while $T_{vx}=\dot T_{xx},~T_{vv}=\dot T_{xv}$ and the initial conditions
at $t=0$ are:
$$
T_{xx}(x_i,v_i,0)=T_{vv}=1,~~~T_{xv}=T_{vx}=0\eqno(5)
$$
$$
\dot T_{xx}(x_i,v_i,0)=\dot T_{vv}=0,~\dot T_{xv}=1,~\dot T_{vx}=-{1\over m}
{d^2V \over dx^2}\Biggl|_{x=x_i}.
$$

Therefore, the dynamics of the distance between the particles $\Delta x$
and the difference of the velocity $\Delta v$ may be represented by the 
non-area-preserving mapping form of the equations of motion
$$
\pmatrix{\Delta x_{i+1}\cr \Delta v_{i+1}}={\bf T}(\alpha;x_i,v_i,\tau_i)
\pmatrix{\Delta x_i\cr \Delta v_i}.\eqno(6)
$$
In general, the intervals between the resets of the velocity $\tau_i$ may
be depending on the number of step $i$.

Further analysis of the model may be based on the general theory of the 
dynamics of classical systems represented as maps [5-9]. Thus, for $\alpha=0$
the Lyapunov exponent is defined as
$$
\lambda={\lim\limits_{N\to\infty}}{1\over N}\sum\limits_{i=1}^N{1\over\tau_i}
\ln\Bigl|{T_{xx}(x_i,v_i,\tau_i)}\Bigr|\eqno(7) 
$$
and may be easily evaluated numerically.

For $\alpha=1$ the map (6) is area-preserving and $\det {\bf T}(1;x_i,v_i,\tau_i)
=1$, while in general $\det {\bf T}=\alpha$, $Tr{\bf T}=T_{xx}+\alpha T_{vv}$ and
the eigenvalues $\mu_{1,2}$ of the ${\bf T}$ matrix are given by equation
$$
\mu^2-\mu Tr{\bf T}+\det {\bf T}=0\eqno(8)
$$
which yields
$$
\mu_{1,2}={1\over 2}\Bigl[T_{xx}+\alpha T_{vv}\mp
\sqrt{(T_{xx}+\alpha T_{vv})^2-4\alpha}\Bigr].\eqno(9)
$$
So, the eigenvalues come in reciprocal pair, $\mu_1\mu_2=\alpha$. For
$(T_{xx}+\alpha T_{vv})^2-4\alpha<0$ the eigenvalues form a complex conjugate
pair with $|\mu_1|=|\mu_2|=\sqrt{\alpha}$, otherwise the eigenvalues are real.

Among the randomly driven systems only dynamics
of the harmonic oscillator with regular time intervals $\tau$ for the reset
of velocity may be described exactly. In this case the ${\bf T}$ matrix is
independent on the starting coordinate $x_i$ and velocity $v_i$, i.e.
$$
{\bf T}=\pmatrix{\cos\omega\tau&(\alpha/\omega)\sin\omega\tau\cr
-\omega\sin\omega\tau&\alpha\cos\omega\tau\cr}.\eqno(10)
$$
If $\alpha<1$ and $\cos\omega\tau\ne\mp 1$ modules of both eigenvalues of the
${\bf T}$ matrix are $|\mu_{1,2}|<1$. So there is a contraction of the distance
in the phase space between the particles in each step. The transition matrix for
$n$ steps ${\bf T}_n={\bf T}^n$ may be calculated by means of diagonalisation of
the matrix (10). The eigenvalues of the ${\bf T}^n$ matrix are $\mu_{1,2}^n$ and
the Lyapunov exponents are simple $\lambda_{1,2}={1\over\tau}\ln|\mu_{1,2}|$.

Generally, the mapping ${\bf T}(\alpha;x_i,v_i,\tau_i)$ in (6) is depending on
the starting coordinates $x_i$ and $v_i$. Therefore, calculation of the
mapping for $n$ steps, ${\bf T}_n={\bf T}(\alpha;x_{i+n-1},v_{i+n-1},\tau_{i+n-1})
\cdot {\bf T}(\alpha;x_{i+n-2},v_{i+n-2},\tau_{i+n-2})~\cdot\cdot\cdot~{\bf
T}(\alpha;x_i,v_i,\tau_i)$,
and of the corresponding eigenvalues are complicated problems. Further we
will evaluate the averaged quantities
$$
\sigma_{1,2}=\bigl\langle{1\over\tau_i}\ln|\mu_{1,2}|\bigr\rangle=
\lim\limits_{N\to\infty}{1\over N}\sum\limits_{i=1}^N{1\over\tau_i}\ln\Bigl|
\mu_{1,2}(x_i,v_i,\tau_i)\Bigr|\eqno(11)
$$
which are analogous of the averaged Lyapunov exponent (7), characterise the
rate of the exponential increase of the separation of the two initially
adjacent points and are related with the $KS$ entropy of the system [5,6].
Comparisons of the threshold values $\tau_c$ from the direct numerical
simulations with those from the criterion 
$$
\sigma_{largest}=0\eqno(12)
$$
indicate to the usefulness of the quantities (11) for analysis of transition 
from nonchaotic to chaotic behavior of the systems.

In Fig. 1 we show typical results of the direct numerical calculations of the
mean-square distance between pairs of identically driven particles in the
Duffing potential $V(x)=x^4-x^2$ vs number of steps for different time
intervals $\tau$ between stops and $\alpha=0$. The averaged characteristic
exponents as functions of time between stops, $\tau$, calculated according to
Eq. (7) for
Duffing and $V(x)=x^4$ potentials are shown in Fig. 2. Regions of $\tau$ with
the positive and negative characteristic exponents correspond respectively to
the chaotic and nonchaotic behaviors. Such a conclusion coincides with the
conclusions from the direct numerical investigation of the systems. In Fig. 3,
we show dependencies of the threshold values for transition from chaotic to 
nonchaotic
behavior of the time $\tau_c$ between resets of the velocity to $v_i^{new}=
\alpha v_i^{old}+v_i^{ran}$ on the parameter $\alpha$ calculated according
to criterion (11)--(12).

For the relatively short time intervals one can expand the position and
velocity of the particles in the powers of time $t$. Using equations of motion
(1) and relations (3) we have expansions for the matrix elements $T_{xx}$ and
$T_{vv}$:
$$
T_{xx}=1-{1\over 2m}{d^2V\over dx_i^2}t^2-{v_i\over 6m}{d^3V\over dx_i^3}t
^3+~\cdot\cdot\cdot
$$
$$
T_{vv}=1-{1\over 2m}{d^2V\over dx_i^2}t^2-{v_i\over 3m}{d^3V\over dx_i^3}t
^3+~\cdot\cdot\cdot\eqno(13)
$$

We can use expressions (13) for the analytical evaluation of the average rate
of contraction of the distance between trajectories for small $\tau$. So, the
result of expansion of eq. (7) according to eq. (13) in powers of $\tau$ is
$$
\lambda\tau=-{1\over 2m}\left<{d^2V\over dx^2_i}\right>\tau^2+\cdots
$$
where the angle brackets denote the average with respect to $x_i$. After the
averagement according to the Boltzmann distribution with $k_BT=m=1$ (instead
of the average with respect to $x_i$ from random trajectories) we have
$\lambda\tau\simeq -2.2\tau^2$ and $\lambda\tau\simeq -2.0\tau^2$ for the
Duffing potential and $V(x)=x^4$ potential, respectively. For $\tau\le 0.5$
these evaluations of the rate coincide with the numerical calculations in Fig.
2.

Moreover, expressions (13) may be used for the description of the systems
with $\alpha\to 1$ and for short time intervals $\tau$. Introducing the
notation $\alpha=1-2\varepsilon~~(\varepsilon\to 0)$ one can have from
equations (9) and (11) the approximate expressions
$$
\mu_{1,2}\simeq 1-\varepsilon\mp\sqrt{T_{xx}+T_{vv}-2+\varepsilon^2}+~\cdot
\cdot\cdot\eqno(14)
$$
$$
\sigma_{1,2}\simeq{1\over\tau}\Bigl[-\varepsilon\mp\bigl\langle(T_{xx}+T_{vv}-
2+\varepsilon^2)^{1/2}\bigr\rangle\Bigr].\eqno(15)
$$
$$
\varepsilon=(1-\alpha)/2\ll 1,~~~\tau\le\tau_c.
$$
The angle brackets in eq. (15) denote the average with respect to $x_i$ and
$v_i$ only if $T_{xx}+T_{vv}-2+\varepsilon^2 > 0$, otherwise $\mu_{1,2}$ are
complex and $\sigma_{1,2}\simeq-\varepsilon/\tau$. The threshold value 
$\tau_c$ of time interval between steps for transition from nonchaotic to
chaotic behavior may be evaluated from the criterion (12).

Using the average according to the Boltzmann and Maxwell distributions
(instead of the average with respect to $x_i$ and $v_i$ from random
trajectories) one can investigate the behavior of the system according to
equations (12)-(15). This yields to the expressions of the threshold time
intervals
$$
\tau_c\simeq 1.44(1-\alpha)/[1+1.6(1-\alpha)],\eqno(16)
$$
$$
\tau_c\simeq (1-\alpha)^{2/3},~~~0.01\le 1-\alpha\ll 1\eqno(17)
$$
for the Duffing and $V(x)=x^4$ potentials, respectively. Here
and in the numerical simulations we set $k_BT=m=1$. It should be noted,
that the "effective temperature" of the system in such simulations is 
$T/(1-\alpha^2)$.

In Fig. 3 the dependencies (16) and (17) are compared with those obtained from 
the numerical calculations according to equations (4), (9), (11) and (12). Note
to the dependence of the law of the approach $\tau_c\to 0$
as $\alpha\to 1$ on the shape of potential.

All such numerical analysis and comparisons with the theoretical results
indicate to the usefulness and fitness of the present theory for the investigation
of transition from chaotic to nonchaotic behavior in randomly driven systems.
Generalization of the theory to more degrees of freedom is straightforward:
for two- or three-dimensional systems equations (1)-(4) and initial conditions
(5) for the ${\bf T}$ matrix are essentially the same, only instead of one
equation (4) we have the systems of two or three coupled equations and dimension
of the ${\bf T}$ matrix is 4 or 6, respectively. Further analysis is rather
similar to that according to equations (6)-(12). The only additional complication
is the diagonalisation of the multidimensional ${\bf T}$ matrix.

In summary, we have developed a theory for transition to nonchaotic behavior
in a Brownian-type motion. Theoretical analysis based on the mapping form of 
equations of motion for the distance between the particles and the difference
of the velocity allows to simplify the problem of investigation of transition
to nonchaotic behavior and results to the expressions for the criteria of the
nonchaotic motion. Theoretical results agree well with the direct numerical
simulations and indicate to the possibilities of generalization of the model,
e.g. to more degrees of freedom, for random values of the time intervals 
between the resets of the velocity and for systems driven by the random forces.

\vskip 2\baselineskip

The research described in this publication was made possible in part by
Grant No. LAA000 from the International Science Foundation.

\vskip\baselineskip
$^*$ Electronic address: kaulakys@itpa.fi.lt

$^\dagger$ Electronic address: vektaris@itpa.fi.lt

\vskip\baselineskip

\begin{enumerate}
\item S. Fahy and D. R. Hamann, Phys. Rev. Lett. {\bf 69}, 761 (1992).
\item D. W. Heermann, {\it Computer Simulation Methods in Theoretical Physics}
(Springer, Berlin, 1990)
\item In the paper [1] the mixing ${\bf v}^{new}=\alpha{\bf v}^{old}+
\beta{\bf v}^{ran}$ has been introduced. However, one can without loss of
generality set $\beta=1$ together with replacement of temperature $T$ by 
$(1-\alpha^2)T$.
\item A. Maritan and J. R. Banavar, Phys. Rev. Lett. {\bf 72}, 1451 (1994).
\item D. Park, {\it Classical Dynamics and Its Quantum Analogues} (Springer,
Berlin, 1990).
\item L. E. Reichl, {\it The Transition to Chaos: In Conservative Classical
Systems: Quantum Manifestations} (Springer, Berlin, 1992).
\item A. J. Lichtenberg and M. A. Lieberman, {\it Regular and Stochastic
Motion} (Springer, New York, 1983).
\item J. D. Meiss, Rev. Mod. Phys. {\bf 64}, 795 (1992).
\item J. M. Ottino {\it et al.}, Science {\bf 257}, 754 (1992).
\end{enumerate}

\newpage

Fig. 1. The mean-square distance $\langle\Delta x^2 \rangle$ between pairs of
identically driven particles in the Duffing potential $V(x)=x^4-x^2$
vs number of steps for different time intervals $\tau$ between stops. 500 
independent simulations with initial conditions from the Gaussian distribution
of variance $\sigma^2=1$ are averaged. The time $\tau$ between stops was
$\tau=1.75$ for the first 300 steps and $\tau=1.3$ for the next steps.

\vskip 2\normalbaselineskip

Fig. 2. Averaged characteristic exponents (Lyapunov exponents 
multiplied by $\tau$) vs the time $\tau$ between stops according to equation
(7) for (a) Duffing potential and (b) $V(x)=x^4$ potential. 

\vskip 2\normalbaselineskip

Fig. 3. Dependencies of the threshold value for transition to chaotic 
behavior of the time $\tau_c$ between resets of the velocity to
$v_i^{new}=\alpha v_i^{old}+v_i^{ran}$ on the parameter $\alpha$ according to
criterion (11) - (12) and according to the analytical limiting expressions
(16) and (17) for (a) Duffing potential and (b) $V(x)=x^4$ potential. 
\end{document}